\documentclass[11pt,twoside]{article}
\usepackage{./asp2014}

\aspSuppressVolSlug
\resetcounters

\begin{document}

\title{Program package for the analysis of high resolution high signal-to-noise stellar spectra}
\author{N. Piskunov$^1$, T. Ryabchikova$^2$, Yu. Pakhomov$^2$, T. Sitnova$^2$, S. Alexeeva$^2$, L. Mashonkina$^2$, and T. Nordlander$^1$}
\affil{$^1$Department of Physics and Astronomy, Uppsala University, Uppsala, Sweden; \email{nikolai.piskunov@physics.uu.se}}
\affil{$^2$Institute od Astronomy RAS, Moscow, Russia; \email{ryabchik@inasan.ru}}

\paperauthor{Piskunov N.}{nikolai.piskunov@physics.uu.se}{}{Uppsala University}{Department of  Physics and Astronomy}{Uppsala}{}{751 20}{Sweden}
\paperauthor{Ryabchikova T.}{ryabchik@inasan.ru}{}{Institute of Astronomy, Russian Academy of Sciences}{}{Moscow}{}{117019}{Russia}
\paperauthor{Pakhomov Yu.}{pakhomov@inasan.ru}{}{Institute of Astronomy, Russian Academy of Sciences}{}{Moscow}{}{117019}{Russia}
\paperauthor{Sitnova T.}{sitnova@inasan.ru}{}{Institute of Astronomy, Russian Academy of Sciences}{}{Moscow}{}{117019}{Russia}
\paperauthor{Alexeeva S.}{alexeeva@inasan.ru}{}{Institute of Astronomy, Russian Academy of Sciences}{}{Moscow}{}{117019}{Russia}
\paperauthor{Mashonkina L.}{lima@inasan.ru}{}{Institute of Astronomy, Russian Academy of Sciences}{}{Moscow}{}{117019}{Russia}
\paperauthor{Nordlander T.}{thomas.nordlander@physics.uu.se}{}{Uppsala University}{Department of  Physics and Astronomy}{Uppsala}{}{751 20}{Sweden}

\begin{abstract}
The program package SME (Spectroscopy Made Easy) \citep{1996AAS..118..595V, 2016arXiv160606073P}, designed to perform an analysis of stellar spectra using spectral fitting techniques, was updated due to adding new functions (isotopic and hyperfine splitting) in VALD and including grids of NLTE calculations for energy levels of few chemical elements. SME allows to derive automatically stellar atmospheric parameters: effective temperature, surface gravity, chemical abundances, radial and rotational velocities, turbulent velocities, taking into account all the effects defining spectral line formation. SME package uses the best grids of stellar atmospheres that allows us to perform spectral analysis with the similar accuracy in wide range of stellar parameters and metallicities -- from dwarfs to giants of BAFGK spectral classes.  
\end{abstract}

\section{Introduction}
The first release of Gaia mission \citep{2016arXiv160904303L} provides us with parallaxes and proper motions for more than 2 million stars. One needs to derive fundamental parameters of these stars -- effective temperature, surface gravity, metallicity -- to restore the stellar population distribution in our Galaxy. The best way to solve this task is the use of fast but robust automatic procedures for stellar spectra analysis because spectroscopy plays the main role in determination of the fundamental parameters. We briefly describe the evolution of the automatic spectral fitting package Spectroscopy Made Easy (SME) developed by \citet{1996AAS..118..595V} and further upgraded by \citet{2016arXiv160606073P}.
 
\section{SME}
SME  computes synthetic spectra and adjusts free parameters ($T_{\rm eff}$ , log~$g$, metallicity, $v_{\rm e}\sin i$, $V_{\rm mic}$, $V_{\rm mac}$, abundances of chemical elements) based on comparison with observations.
SME solver consists of the IDL routines for preparing spectral synthesis and performing optimization, and external library for synthetic spectrum calculations. The external library (SYNTH code) is written in C++ and Fortran.
SME spectral synthesis includes molecular and ionization equilibrium solver EOS, continuous opacity package CONTOP, line opacity package LINEOP and radiative transfer (RT) solver. EOS has partition functions for up to 6 ionization stages of the first 99 atoms in the periodic table and for 257 molecules (up to four atoms) fitted over the range from 10~K to 8000~K \citep{2016AA...588A..96B}.
EOS is using rather unique solving strategy.
SME is working with the observations in ASCII or FITS formats.
The format of input linelist is one of the output formats of VALD 'Extract Stellar' request \citep{1999AAS..138..119K}. Short format is used when SME is working in LTE (local thermodynamic equilibrium) regime, while for NLTE (non-local thermodynamic equilibrium) analysis a long format is needed. The current version of SME has model libraries of Kurucz' (1993) models,  the latest version of MARCS models \citep{2008AA...486..951G}, and a grid of LLmodels \citep{2004AA...428..993S} calculated for microturbulent velocity of 2~km\,s$^{-1}$. Both plane-parallel and spherical MARCS models are available.

In general, SME provides reliable estimates of the atmospheric parameters for dwarf and giant stars. For dwarfs it is justified in the papers by \citet{2016MNRAS.456.1221R}, and \citet{2015ASPC..494..308R}. For two cool giant stars the SME results are compared with other published determinations in Table~\ref{giant}. 

\begin{table}
\vskip -10pt
\begin{center}
\caption{Atmospheric parameters of the giant stars derived by different methods. \label{giant}}
\begin{tabular}{rcccl}
\tableline\noalign{\smallskip}
 HD & $T_{\rm eff}$        & log~$g$      &  [Fe/H]        &Reference  \\
\tableline \noalign{\smallskip}
 4306    & 4944(122) &1.95(48)&$-$2.80(10)& This paper (SME)\\
 4306    & 4800(~40) &1.70(06)&$-$2.92(04)& \citet{2000AJ....120.1841F}\\
 4306    & 5000(100) &2.10(30)&$-$2.52(15)& \citet{2001AA...370..951M}\\
\tableline\noalign{\smallskip}
74387    & 4833(~40) &2.41(16)&$-$0.29(05)& This paper (SME)\\                   
74387    & 4840(100) &2.43(10)&$-$0.18(05)& \citet{2009ARep...53..685P} \\                   
\tableline 
\end{tabular}
\end{center}
\vskip -25pt
\end{table}

\section{NLTE grid calculations}
SME allows us to derive not only atmospheric parameters but element abundances as well. Abundances are derived as average over all observed ions. Original SME version \citep{1996AAS..118..595V} was working with the spectral synthesis under LTE assumption. Some lines of important chemical elements were excluded from fitting procedure because of NLTE effects. The current SME version includes a possibility to use precalculated NLTE data in fitting. These data consist of departure coefficients $b$ (a ratio of NLTE to LTE level populations) for energy levels of an element. Calculations are performed for grids of MARCS plane-parallel models with the microturbulent velocity of 1~km\,s$^{-1}$\ using DETAIL code \citep{detail}. At present we have NLTE data grids for elements O~I, Na~I, Ca~I-II, and Ba~II.

\underline{Oxygen}. NLTE oxygen calculations were performed using model atom from \citet{2000AA...359.1085P} that includes 51 levels of O~I and ground state of O~II. Electron-impact exitation data from \citet{2007AA...462..781B} were employed.

\underline{Sodium}  17 levels with n$\leq$7, l$\leq$5 of Na~I and ground state of Na~II represent Na model atom \citep{2014AstL...40..406A}. Cross-sections for electron-impact collisions were extracted from \citet{2008ADNDT..94..981I} while rate coefficients for inelastic collisions with hydrogen atoms were taken from \citet{2010AA...519A..20B}.  

\underline{Calcium}  Calcium model atom includes 63 levels of Ca~I, 37 levels of Ca~II and ground state of Ca~III \citep{2007AA...461..261M}. 

\underline{Barium}. For barium we used model atom developed by \citet{1999AA...343..519M}. It includes 35 levels of Ba~II and ground state of Ba~III.

Search for final atmospheric parameters and abundances is carried out by interpolation between the nodes in model and NLTE data grids as described in details in \citet{2016arXiv160606073P}. 

Comparison of the model atmosphere structure and NLTE departure coefficients $b$ obtained by direct calculations and by interpolation process shows fairly good agreement. Even in the cases of complex dependence of $b$-factors on atmospheric depth when interpolation smooths the curves out, it does not influence too much the emergent line flux (see Fig.~\ref{NLTE-O}). The difference in the equivalent width of O~I 7771.94~\AA\ line does not exceed 0.2\% that translates to 0.02~dex of abundance difference.
     
\articlefigure{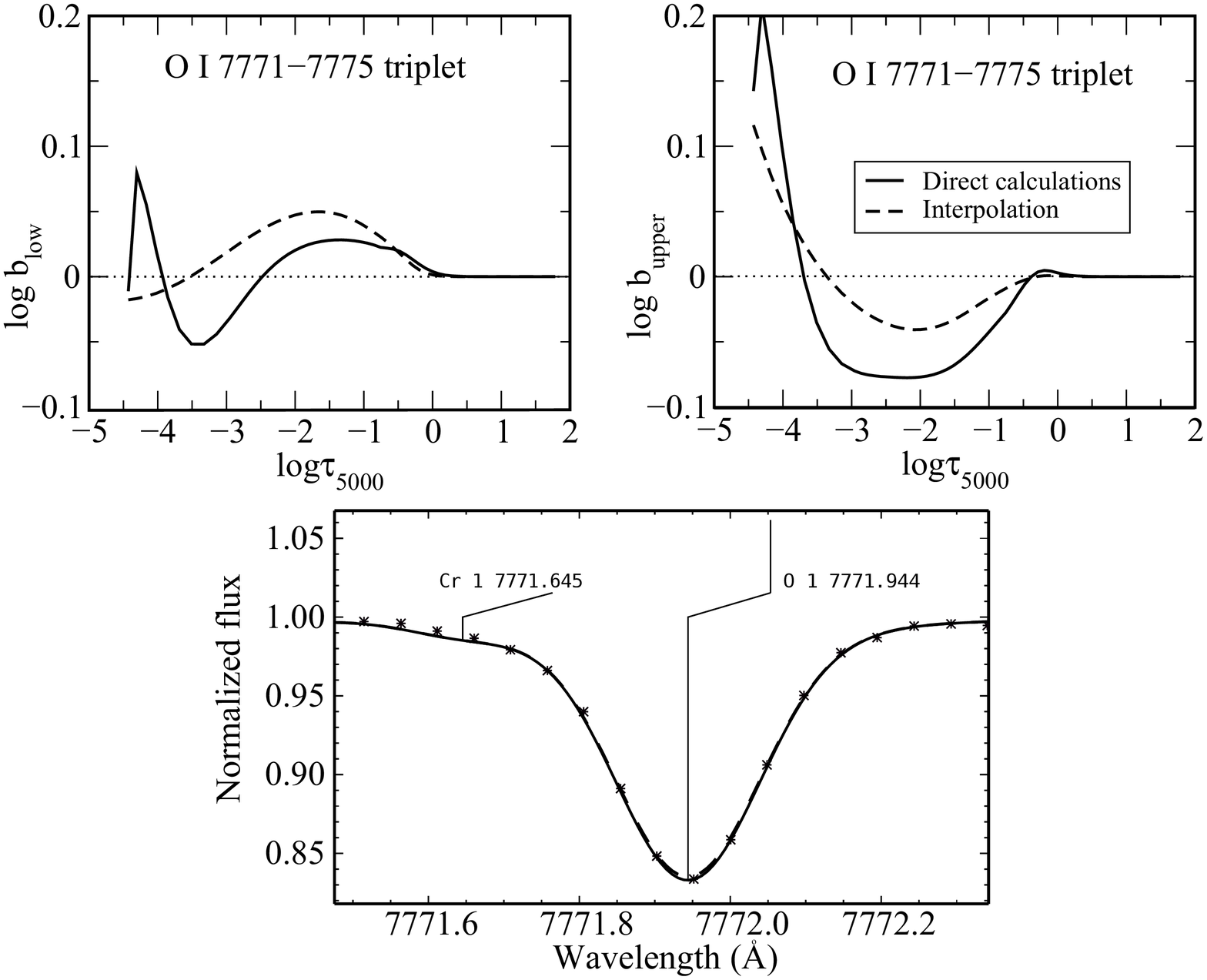}{NLTE-O}{\textit{Top panels}: comparison between $b$-factors derived by direct NLTE calculations (\textit{full lines}) and by interpolation (\textit{dashed lines}) for lower (left) and upper (right) energy levels of O~I IR-transitions. Calculations are made for HD~69830 ($T_{\rm eff}$=5422~K, log~$g$=4.47, metallicity=-0.04) \textit{Bottom panel}: comparison between line profiles of O~I 7771.94~\AA\ line calculated with $b$-departure coefficients described above. Observations are shown by \textit{asterisks}.}

We compared abundances determined with SME and by separate fitting of carefully chosen set of spectral lines in the atmosphere of cool dwarf HD~69830. SME-derived parameters were taken from \citet{2016MNRAS.456.1221R} - ESPaDONs spectrum. The procedure of line-by-line abundance determination is given in the same paper as well. Note, that in both cases abundances were derived using different sets of spectral lines. This comparison shows fairly good agreement of both sets of abundances within the typical errors of best solar abundance determinations, $\pm$0.04~dex.   

\articlefigure[width=0.7\textwidth]{SME-LbL_HD69830.eps}{Abun-comp}{Comparison between abundances derived for HD~69830 with SME and by separate fitting of carefully chosen set of spectral lines. Elements analysed under NLTE assumption are marked by open circles. Errors are shown for SME analysis.}

\acknowledgements This work was partially supported by the grant of the Leading School No\,9951.2016.2 and by the 
Russian Foundation for Basic Research (grant 15-02-06046).

\bibliography{references_2016a}  

\begin{thebibliography}{}
\expandafter\ifx\csname natexlab\endcsname\relax\def\natexlab#1{#1}\fi
\expandafter\ifx\csname url\endcsname\relax
  \def\url#1{\texttt{#1}}\fi
\expandafter\ifx\csname urlprefix\endcsname\relax\def\urlprefix{URL }\fi
\providecommand{\eprint}[2][]{\url{#2}}

\bibitem[{{Alexeeva} et~al.(2014){Alexeeva}, {Pakhomov}, \&
  {Mashonkina}}]{2014AstL...40..406A}
{Alexeeva}, S.~A., {Pakhomov}, Y.~V., \& {Mashonkina}, L.~I. 2014, Astronomy
  Letters, 40, 406

\bibitem[{{Barklem}(2007)}]{2007AA...462..781B}
{Barklem}, P.~S. 2007, \aap, 462, 781

\bibitem[{{Barklem} et~al.(2010){Barklem}, {Belyaev}, {Dickinson}, \&
  {Gad{\'e}a}}]{2010AA...519A..20B}
{Barklem}, P.~S., {Belyaev}, A.~K., {Dickinson}, A.~S., \& {Gad{\'e}a}, F.~X.
  2010, \aap, 519, A20

\bibitem[{{Barklem} \& {Collet}(2016)}]{2016AA...588A..96B}
{Barklem}, P.~S., \& {Collet}, R. 2016, \aap, 588, A96

\bibitem[{{Butler} \& {Giddings}(1985)}]{detail}
{Butler}, K., \& {Giddings}, J. 1985, Newsletter on the analysis of
  astronomical spectra, No. 9, University of London

\bibitem[{{Fulbright}(2000)}]{2000AJ....120.1841F}
{Fulbright}, J.~P. 2000, \aj, 120, 1841

\bibitem[{{Gustafsson} et~al.(2008){Gustafsson}, {Edvardsson}, {Eriksson},
  {J{\o}rgensen}, {Nordlund}, \& {Plez}}]{2008AA...486..951G}
{Gustafsson}, B., {Edvardsson}, B., {Eriksson}, K., {J{\o}rgensen}, U.~G.,
  {Nordlund}, {\AA}., \& {Plez}, B. 2008, \aap, 486, 951

\bibitem[{{Igenbergs} et~al.(2008){Igenbergs}, {Schweinzer}, {Bray}, {Bridi},
  \& {Aumayr}}]{2008ADNDT..94..981I}
{Igenbergs}, K., {Schweinzer}, J., {Bray}, I., {Bridi}, D., \& {Aumayr}, F.
  2008, Atomic Data and Nuclear Data Tables, 94, 981

\bibitem[{{Kupka} et~al.(1999){Kupka}, {Piskunov}, {Ryabchikova}, {Stempels},
  \& {Weiss}}]{1999AAS..138..119K}
{Kupka}, F., {Piskunov}, N., {Ryabchikova}, T.~A., {Stempels}, H.~C., \&
  {Weiss}, W.~W. 1999, \aaps, 138, 119

\bibitem[{{Lindegren}(2016)}]{2016arXiv160904303L}
{Lindegren}, L. et~al. 2016, ArXiv e-prints. \eprint{1609.04303}

\bibitem[{{Mashonkina} et~al.(1999){Mashonkina}, {Gehren}, \&
  {Bikmaev}}]{1999AA...343..519M}
{Mashonkina}, L., {Gehren}, T., \& {Bikmaev}, I. 1999, \aap, 343, 519

\bibitem[{{Mashonkina} et~al.(2007){Mashonkina}, {Korn}, \&
  {Przybilla}}]{2007AA...461..261M}
{Mashonkina}, L., {Korn}, A.~J., \& {Przybilla}, N. 2007, \aap, 461, 261

\bibitem[{{Mishenina} \& {Kovtyukh}(2001)}]{2001AA...370..951M}
{Mishenina}, T.~V., \& {Kovtyukh}, V.~V. 2001, \aap, 370, 951

\bibitem[{{Pakhomov} et~al.(2009){Pakhomov}, {Antipova}, {Boyarchuk}, {Zhao},
  \& {Liang}}]{2009ARep...53..685P}
{Pakhomov}, Y.~V., {Antipova}, L.~I., {Boyarchuk}, A.~A., {Zhao}, G., \&
  {Liang}, Y. 2009, Astronomy Reports, 53, 685

\bibitem[{{Piskunov} \& {Valenti}(2016)}]{2016arXiv160606073P}
{Piskunov}, N., \& {Valenti}, J.~A. 2016, ArXiv e-prints. \eprint{1606.06073}

\bibitem[{{Przybilla} et~al.(2000){Przybilla}, {Butler}, {Becker}, {Kudritzki},
  \& {Venn}}]{2000AA...359.1085P}
{Przybilla}, N., {Butler}, K., {Becker}, S.~R., {Kudritzki}, R.~P., \& {Venn},
  K.~A. 2000, \aap, 359, 1085

\bibitem[{{Ryabchikova} et~al.(2016){Ryabchikova}, {Piskunov}, {Pakhomov},
  {Tsymbal}, {Titarenko}, {Sitnova}, {Alexeeva}, {Fossati}, \&
  {Mashonkina}}]{2016MNRAS.456.1221R}
{Ryabchikova}, T., {Piskunov}, N., {Pakhomov}, Y., {Tsymbal}, V., {Titarenko},
  A., {Sitnova}, T., {Alexeeva}, S., {Fossati}, L., \& {Mashonkina}, L. 2016,
  \mnras, 456, 1221

\bibitem[{{Ryabchikova} et~al.(2015){Ryabchikova}, {Piskunov}, \&
  {Shulyak}}]{2015ASPC..494..308R}
{Ryabchikova}, T., {Piskunov}, N., \& {Shulyak}, D. 2015, in Physics and
  Evolution of Magnetic and Related Stars, edited by Y.~Y. {Balega}, I.~I.
  {Romanyuk}, \& D.~O. {Kudryavtsev}, vol. 494 of Astronomical Society of the
  Pacific Conference Series, 308

\bibitem[{{Shulyak} et~al.(2004){Shulyak}, {Tsymbal}, {Ryabchikova},
  {St{\"u}tz}, \& {Weiss}}]{2004AA...428..993S}
{Shulyak}, D., {Tsymbal}, V., {Ryabchikova}, T., {St{\"u}tz}, C., \& {Weiss},
  W.~W. 2004, \aap, 428, 993

\bibitem[{{Valenti} \& {Piskunov}(1996)}]{1996AAS..118..595V}
{Valenti}, J.~A., \& {Piskunov}, N. 1996, \aaps, 118, 595

\end{thebibliography}

\end{document}